\documentclass[aps,prapplied,reprint,groupedaddress]{revtex4-2}
\usepackage{graphicx}
\usepackage{amsmath}
\usepackage{titlesec}
\usepackage{booktabs}
\usepackage{lipsum}
\usepackage{multirow}
\usepackage[hidelinks]{hyperref}

\titleformat{\section}
  {\normalfont\bfseries\filcenter\uppercase}  
  {\thesection}{1em}{}

\newcommand{\authormark}[1]{\textsuperscript{#1}}

\begin{document}
\preprint{}

\title{Improving key rates by tighter information reconciliation leakage estimation for quantum key distribution}
\thanks{H.-K. M. and B. Y. contributed equally to this work.}

\author{Hao-Kun Mao\authormark{1, *}, Bo Yang\authormark{1, *}, Yu-Cheng Qiao\authormark{2}, Bing-Ze Yan\authormark{1}, Qiang Zhao\authormark{3}, Bing-Jie Xu\authormark{4} and Qiong Li\authormark{1,}}
\email{Contact author: qiongli@hit.edu.cn}
\affiliation{\authormark{1}School of Cyberspace Science, Faculty of Computing, Harbin Institute of Technology, Harbin 150080, China\\
\authormark{2}Guangxi Key Lab Cryptography and Information Security, Guilin University of Electronic Technology, Guilin 541004, Guangxi, China\\
\authormark{3}School of Information Engineering, Zhejiang Ocean University, Zhoushan 316022, China\\
\authormark{4}Science and Technology on Security Communication Laboratory, Institute of Southwestern Communication, Chengdu 610041, China}


\date{\today}

\begin{abstract}
Previous research has aimed to precisely estimate information leakage  to improve the secure key rate (SKR) and maximum transmission distance in quantum key distribution (QKD). However, existing methods repeatedly considerd the information of the multi-photon pulses known to Eve before and after information reconciliation, resulting in an overestimation of the leakage amount. We propose a novel approach that considers the quantum part's effect on post-processing, providing a more accurate estimation of information reconciliation leakage to improve the SKR. Theoretical analysis shows that our method more accurately estimates the information reconciliation leakage, significantly improving the SKR at any distance as well as the maximum transmission distance. It is worth mentioning that previous studies treat information leakage of the error correction as Shannon bound $Nh(e)$, and our method can estimate it more tightly. Simulation results for decoy-BB84 and measurement-device-independent (MDI) protocols using Cascade are consistent with the theoretical analysis. The farther the transmission distance, the greater the growth rate of SKR. When the error rate is 33\%, compared with the original method, the SKR growth rate of decoy-BB84 at 100KM is 100.4\%, and the transmission distance of MDI increases by 22KM.
\end{abstract}


\maketitle

\section{Introduction}
\label{sec:1}
Quantum key distribution (QKD) constitutes a promising solution for distributing unconditionally secure keys between two remote parties, such as Alice and Bob, even in the presence of an eavesdropper, Eve. Since the first QKD protocol, known as the BB84 protocol, was proposed by C. H. Bennett and G. Brassard in  1984 \cite{bennett1984quantum}, security analysis has been a central focus of QKD studies. Although the ideal BB84 protocol has been proven to be unconditionally secure \cite{shor2000simple}, practical implementations may introduce security vulnerabilities due to imperfect devices, which can threaten the security of a QKD system. One significant vulnerability arises from the photon-number-splitting (PNS) attack \cite{norbertlutkenhaus2000security, brassard2000security, norbertlutkenhaus2002quantum, ashkenazy2024photon, sixto2023secret}. In this attack, Eve can exploit imperfections in photon sources to obtain complete information about each multi-photon pulse without altering the quantum bit error rate (QBER). 

To address this challenge, Gottesman-Lo-Lutkenhaus-Preskill (GLLP) \cite{gottesman2004security} proved the unconditional security of practical QKD systems with imperfect devices, providing a robust theoretical foundation for ensuring the security of practical QKD systems and mitigating the vulnerabilities introduced by imperfect devices. In general, the SKR formula based on the GLLP theory contains two main parts: the secure amount of single-photon information which is completely unknown to Eve, and the reconciliation information leakage which is used to correct the errors. The former part has attracted sustained attention due to its potential in improving SKR, introducing many theories, such as the finite size effect analysis \cite{ma2005practical, tomamichel2012tight, scarani2008quantum, zhou2016making, lim2014concise, rusca2018finitekey, fan-yuan2021optimizing}, and also introducing numerical methods by considering realistic imperfections and providing reliable security proofs \cite{wang2021numerical, tupkary2023using}. In contrast, the reconciliation information leakage, which is used to correct the errors, has largely been overlooked since it is commonly believed to be constrained by the Shannon limit. Nevertheless, we believe that directly treating this information leakage as the Shannon limit is too conservative. In this paper, the information leakage of this part can be further decreased by considering the effect of multi-photon pulses to the reconciliation, thereby improving the SKR of QKD.

We first assume that random variables $A$ and $B$ represent the sequences of Alice and Bob to be reconciled of length $N$, respectively. According to the noiseless coding theorem, the lower bound of the exchanged information $L_{all}$ for reliable reconciliation can be calculated by the conditional entropy $H(A|B)$ \cite{slepian1973noiseless}. In a DV-QKD system, $H(A|B)$ can be written as $Nh(e)$, where $e$ is referred to as QBER and the binary Shannon entropy $h(e) =  - e{\log_2}(e) - (1 - e){\log_2}(1 - e)$. Nevertheless, in our view, the actual information leakage of reconciliation $L_{actual}$ is not equal to, but less than $L_{all}$ in practical DV-QKD systems. We notice that in the exchanged information $L_{all}$, the information leakage caused only by multi-photon pulses $L_M$ is supposed to completely known by Eve before reconciliation and unable to provide any extra information for Eve after reconciliation. Thus, $L_M$ should not be subtracted again after reconciliation from the candidate secure keys. In other words, $L_M$ can be considered to have no impact on SKR. As long as $L_{actual} = L_{all} - L_M$ is eliminated, the candidate keys can be guaranteed to be secure. In this way, the unnecessary information leakage $L_{M}$ can be avoided, thus improving the SKR. Here, we just give a brief introduction to the main idea of our approach while the details will be discussed in Sec. \ref{sec:2}.

The rest of this paper is organized as follows. Sec. \ref{sec:back} provides the background information required. Sec. \ref{sec:2} presents the main idea of our approach and the formulas for calculating information leakage for two typical protocols are given in Sec. \ref{sec:3}. The performances of our approach are reported and analyzed through numerical simulations in Sec. \ref{sec:4} and the concluding remarks are made in Sec. \ref{sec:5}.
\section{BACKGROUND}
\label{sec:back}
\subsection{Photon-number-splitting attack}
In the Shor-Preskill security analysis of the ideal BB84 protocol, it is assumed that the QKD system uses an ideal single-photon source \cite{shor2000simple}. However, in practical BB84 QKD systems, a weak coherent pulse is used as the light source, which has a certain probability of emitting multi-photon signals. Eve can exploit this security vulnerability to perform a PNS attack, fully extracting the information encoded in multi-photon signals without causing any error.

The process of PNS attack is as follows:

\textbf{Step 1} Non-Demolition Measurement and Blocking. Eve performs a quantum non-demolition (QND) measurement on the light pulses emitted by Alice to detect the number of photons in each pulse. In this step, Eve does not disturb the encoded information within the pulses and only measures the photon count. If the pulse contains only one photon, Eve blocks the single-photon signal with probability \( p \) based on the total transmission loss \( \eta \) in the QKD system, where \( \eta \) represents the total loss in the QKD system; if the pulse contains multiple photons, Eve proceeds to the next step.

\textbf{Step 2} Photon Splitting and Storage. When Eve detects multiple photons in the pulse, she intercepts one photon and stores it in a quantum memory. The remaining photons continue to be sent to Bob. During this process, Eve saves part of the pulse information through quantum memory technology, without affecting Bob's detection rate, thus avoiding detection by Alice and Bob.

\textbf{Step 3} Measurement and Key Extraction. After Alice and Bob complete basis reconciliation and disclose information, Eve performs a measurement on the stored photons based on the disclosed basis information to obtain the encoded information using the correct basis. This approach enables Eve to steal part of the key information without significantly disturbing the QKD system.

By selectively intercepting and storing multi-photon pulses, photon-number-splitting attacks can compromise key information while preserving the integrity and consistency of data transmission in QKD systems, thus circumventing certain detection mechanisms. This type of attack challenges the security framework of QKD systems and encourages researchers to explore practical security analysis of QKD.
\subsection{Decoy-state protocol}
The PNS attack significantly limits the secure key rate and transmission distance of practical QKD systems. In the BB84 protocol, the secure key rate scales as \( O(\eta^2) \). To counter the PNS attack, theorists proposed the decoy-state protocol, which improves the secure key rate scaling to \( O(\eta) \) \cite{scarani2009security}.

The core security mechanism of the decoy-state protocol lies in detecting potential Photon Number Splitting (PNS) attacks by using optical pulses of different intensities (i.e., signal states and decoy states). The sender, Alice, randomly selects different pulse intensities to send signal and decoy states, while the receiver, Bob, performs independent statistical analysis on each state. 

The key aspects of the decoy-state protocol are as follows:

\textbf{(1) }Distinguishing Between Signal and Decoy States. In the decoy-state protocol, the presence of decoy states prevents Eve from distinguishing between signal and decoy states. Consequently, Eve cannot employ different strategies for signal and decoy states during eavesdropping. Through statistical analysis, Alice and Bob can detect anomalies in measurement results caused by eavesdropping, thus identifying potential PNS attacks.
	
\textbf{(2) }Detection of Statistical Deviations. The security of the decoy-state protocol relies on the statistical analysis of results from signal and decoy states. If the system is secure, the measurement results for pulses of different intensities should match the expected statistical characteristics. However, if Eve performs a PNS attack, these statistical characteristics will deviate. Alice and Bob can identify and quantify the information leakage risk caused by eavesdropping through these deviations, ensuring the security of the final key.

The decoy-state protocol enables QKD systems to detect and defend against PNS attacks without relying on ideal single-photon sources. This approach allows QKD systems to maintain a high SKR and transmission distance under practical physical conditions, thereby significantly enhancing the practical security of QKD systems.   
\subsection{Cascade}
The core idea of Cascade is to discover errors through parity check and backtracking, and correct them using binary operations (\textbf{BISECT}). To ensure smooth error correction, Cascade performs multiple rounds of error correction, with each round requiring partitioning of Alice's and Bob's data frames \cite{martinez-mateo2015demystifying}.

The parity check operation is relatively simple. Assuming the data block length is \( k \), the corresponding data blocks for Alice and Bob are \( a = \left( a_{1}, a_{2}, \dots, a_{k} \right) \) and \( b = \left( b_{1}, b_{2}, \dots, b_{k} \right) \), respectively. Parity check requires first calculating parity values \( s_{a} = a_{1} \oplus a_{2} \oplus \dots \oplus a_{k} \) and \( s_{b} = b_{1} \oplus b_{2} \oplus \dots \oplus b_{k} \) (where \( \oplus \) is the "XOR operation" i.e., modulo 2 summation) and comparing \( s_{a} \) and \( s_{b} \) through interaction. If \( s_{a} \neq s_{b} \), blocks \( a \) and \( b \) contain an odd number of errors, and one error can be corrected using the \textbf{BISECT} operation; if they match, any undetected errors will be corrected in subsequent rounds. This is why Cascade needs to perform multiple rounds of error correction.

The workflow of the \textbf{BISECT} operation is as follows:

\textbf{Step 1} If \( |a| \neq 1 \) and \( |b| \neq 1 \), proceed to Step 2; otherwise, Bob inverts \( b \), ending the BISECT process.

\textbf{Step 2} Alice splits \( a \) into \( a = a_1 a_2 \), with \( |a_1| = \left\lfloor \frac{|a|}{2} \right\rfloor \) and \( |a_2| = \left\lceil \frac{|a|}{2} \right\rceil \). Alice sends the parity \( s_{a_1} \) of \( a_1 \) to Bob (alternatively, \( s_{a_2} \) could be sent). Bob performs the same operations for his block.

\textbf{Step 3} If \( s_{a_1} = s_{b_1} \), it follows that \( s_{a_2} \neq s_{b_2} \) due to \( s_a \neq s_b \), meaning there is an odd number of errors in \( a_2 \) and \( b_2 \). Set \( a = a_2 \) and \( b = b_2 \); otherwise, set \( a = a_1 \) and \( b = b_1 \). Return to Step 1 to continue the BISECT process.

Another key operation in Cascade is backtracking, which uses the data relationships across rounds to correct errors with minimal information leakage. Assuming that error correction is performed on the $i\left( i\geq 2\right) $-th round, it indicates that the data blocks in the previous $i-1$ rounds all contain an even number of errors or no errors; If a new error code is discovered in this round, the data block from the previous $i-1$ rounds must contain an odd number of errors. At this point, not only can the overhead of discovering errors through parity check be saved by 1-bit, but the length of the first $i-1$ data blocks is usually smaller, and the amount of leaked information in BISECT operation will also be less.

The complete Cascade protocol workflow is as follows:

\textbf{Step 1} In the first round \( t = 1 \), Alice selects a random permutation \( f_1: [1..n] \rightarrow [1..n] \) to shuffle her sequence \( A \) of length \( n \) and determines the grouping of data blocks of length \( k_1 \). The position set \( K_v^1 = \{ i \mid \lceil f_1(i)/k_1 \rceil = v \} \) indicates the \( v \)-th data block’s positions post-shuffling. Bob performs the same operation. Both parties then exchange and compare parity positions, and blocks with mismatched parity undergo \textbf{BISECT}. Data block information is recorded as \( S_1 = \{ K_v^1 \mid 1 \leq v \leq \lceil n/k_1 \rceil \} \) and and update \( t = 2 \).

\textbf{Step 2} In round \( t \), Alice and Bob choose a new permutation \( f_t: [1..n] \rightarrow [1..n] \) and block size \( k_t \) for shuffling. The initial positions of blocks are stored as \( S_t = \{ K_v^t \mid 1 \leq v \leq \lceil n/k_t \rceil \} \), with blocks permuted sequentially, setting \( j = 1 \);

\textbf{Step 2-1} Both calculate and compare the \( j \)-th block’s parity. Discrepancies trigger the \textbf{BISECT} process to locate errors, recording the error position as \( l_1 \), and moving to \textbf{Step 2-2}; otherwise, proceed to \textbf{Step 2-3};

\textbf{Step 2-2} (Backtracking) Let \( P_1 \) be blocks from rounds \( 1 \sim (t-1) \) containing \( l_1 \), i.e., \( P_1 = \{ K \mid K \in \bigcup_{a=1}^{t-1} S_a \land l_1 \in K \} \), which contains an odd number of errors except for \( l_1 \). Update the global set \( \beta = P_1 \), and at this time, both parties select the shortest data block from \( \beta \) for the \textbf{BISECT} operation (this data block is no longer in the set \( \beta \)). Assume a new error code position \( l_2 \) is discovered. Similarly, let the set \( P_2 \) represent the data blocks recorded from rounds \( 1 \sim (t-1) \) that contain error position \( l_2 \), i.e., \( P_2 = \{ K \mid K \in \bigcup_{a=1}^{t-1} S_a \land l_2 \in K \} \). At this point, add \( P_2 \) to the global set \( \beta \), i.e., \( \beta = (\beta \cup P_2) \setminus (\beta \cap P_2) \). Note that if \( \beta \cap P_2 \neq \emptyset \), it indicates that the current error code has backtracked to a data block containing an odd number of error codes, so this data block needs to be removed. Repeat the above process until \( \beta = \emptyset \);

\textbf{Step 2-3} If \( j < \lceil n/k_1 \rceil \), it means there are still unchecked data blocks. Set \( j \leftarrow j + 1 \) and proceed to \textbf{Step 2-1}; otherwise, proceed to \textbf{Step 3}.

\textbf{Step 3} If \( t \) exceeds the maximum rounds \( t_{max} \), the protocol completes; otherwise, set \( t \leftarrow t + 1 \) and jump to Step 2.

This outlines the workflow of the Cascade protocol.

\begin{figure*}
	\includegraphics[width=\textwidth]{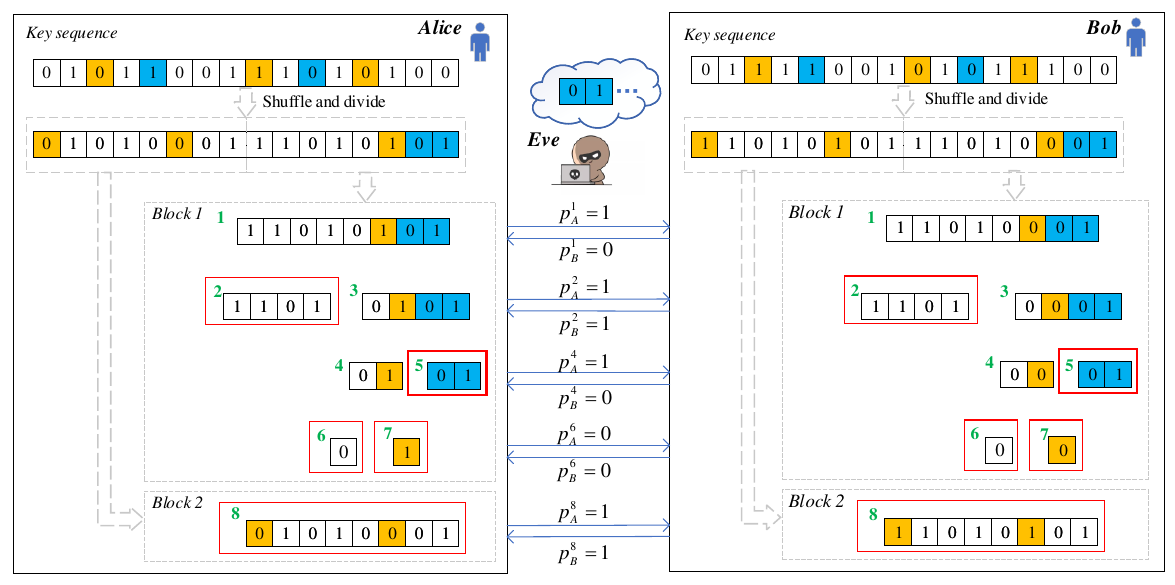}
	\caption{\label{fig:1}\textbf{A schematic workflow of Cascade protocol.} The keys that are inconsistent in two parties and generated from single-photon and multi-photon pulses are filled with orange and blue, respectively. All the blocks (sub-blocks) generated during the error correction process are numbered in green, and $p_A^i$, $p_B^i$ represent the parity (i.e., the sum modulo 2 of all bits) of the ith block belonging to Alice and Bob, respectively. }
\end{figure*}

\section{Finer estimation of information leakage reconciliation}
\label{sec:2}
In this section, we only take Cascade reconciliation protocol as an example to further elaborate on the main idea of our approach. The ideas of this method can also be applied to other information reconciliation protocols. As illustrated in Fig. (\ref{fig:1}), assume that the two inconsistent keys generated by single-photon and multi-photon pulses are filled with orange and blue, respectively, and that Alice and Bob are unaware of this information. The key sequences of Alice and Bob are first shuffled and divided into two blocks with length of $8$. Then, the parities of “Block 1” and “Block 2” in both parties are exchanged and compared simultaneously.

For “Block 1” with $p_A^1 \ne p_B^1$, a binary search operation (i.e., recursively split into two and check the parity of the first half) is performed to correct one error bit. Note that there is no need to exchange the parity of the second half since it can be easily obtained. For example, $p_A^3$ can be calculated by $p_A^1 \oplus p_A^2$. After four rounds of communications, the error is eventually discovered in the 7th sub-block. Through the exchanged parity information, Eve gains more information about the key sequences. For instance, the number of possible combinations of Alice’s data in the 2th sub-block decreases from 16 to 8 with the help of $p_A^2=1$. Overall, Eve obtains 1-bit information from each sub-block boxed in red and the total exchanged information of “Block 1” is 4-bit. 

However, when considering the effect of the quantum part on the post-processing part, the actual amount of information by Eve from reconciliation changes. It is worth mentioning that we considered blocks consisting entirely of multi-photon pulses in this study. Therefore, let us now focus on the special 5th sub-block which is composed of two bits generated from multi-photon pulses. Before reconciliation, Eve has already obtained all the information of these two bits through PNS attack. Though the parity is leaked to Eve during reconciliation, Eve cannot acquire any extra useful information from this. In other words, the parity information has been known by Eve even if the parity is not exchanged. In this way, only the parity information obtained from the 2th, 6th and 7th sub-blocks are useful for Eve. Thus, the actual amount information obtained by Eve is only 3-bit, even though 4-bit information has been exchanged during reconciliation.  

We do nothing with “Block 2” with matched parity that $p_A^8=p_B^8$, even if this block still contains even number of undetected errors which will be corrected in the subsequent passes. After several iterative passes of error correction (i.e. restarting shuffling and binary search operations in each pass), the probability of no remaining errors is rather high, signifying a successful error correction. In addition, Cascade involves backtracking operations to benefit the binary search operation, thereby reducing the information leakage. 

The above analysis suggests that the information reconciliation leakage of a sub-block whose bits are all generated by multi-photon pulses is $0$ bit. Thus, the actual amount of information leakage $\operatorname{leak}_{\text {all }}^{\mathrm{EC}}$ to Eve can be obtained by calculating the number of such blocks. We can adjust the reconciliation process to ensure that no redundant information is subtracted. This adjustment not only improves the SKR but also strengthens the system's resilience against eavesdropping. Thus, the precise calculation of $\operatorname{leak}_{\text {all }}^{\mathrm{EC}}$ is an essential step in refining the security protocols of QKD, ensuring that the system remains robust even under practical conditions with imperfect devices.
\section{New information leakage estimation for two typical protocols}
\label{sec:3}
\subsection{Theoretical analysis }
The known formula for the asymptotic SKR is given by the difference between two information-theoretic quantities, related to privacy amplification (PA) and error correction (EC), respectively \cite{devetak2005distillation}. The privacy amplification term accounts for the information that needs to be subtracted from the raw key to ensure that the final key remains secure from any potential eavesdropping by an adversary, commonly referred to as Eve. The error correction term is necessary to reconcile discrepancies between the keys held by the legitimate parties, Alice and Bob, due to noise in the quantum channel and imperfections in the measurement devices. 

Let $R$ represents the SKR in a practical DV-QKD system, the expression for SKR of these two terms (PA and EC) for each signal is as follows  \cite{tupkary2023using, nickykaihongli2020improving}
\begin{equation}
	\label{eq:1}
		R = p_{\text {pass }} \cdot \left(\min _{\rho \in \mathbf{S}} \widehat{f}(\rho)-\operatorname{leak}_{\text {all }}^{\mathrm{EC}}\right),\\
\end{equation}
where $p_{\text {pass }}$ is probability of signal passing through sifting.  The sifting process is a critical step in which Alice and Bob compare a subset of their data over a public channel to detect and discard any signals that do not match, thereby enhancing the security of the remaining key. $\mathbf{S}$ is the set of all density matrices that satisfy the joint statistic of Alice and Bob. $\min _{\rho \in \mathbf{S}} \widehat{f}(\rho)$ represents the minimum relative entropy of some two classical-quantum joint states in a quantum and classical joint system (where $\rho$ is the density matrix). The $\operatorname{leak}_{\text {all}}^{\mathrm{EC}}$ is the number of bits used in the error correction process, per bit of the raw key, and the meanings of the other parameters above are in Ref. \cite{tupkary2023using}. 

Recent studies have expanded on these concepts by incorporating various techniques to improve the practical implementation of QKD systems. For example, Tupkary and Lütkenhaus \cite{tupkary2023using} have investigated the integration of error location information into the initial phase of the Cascade protocol to achieve a more stringent lower bound for SKR. Li et al. \cite{nickykaihongli2020improving} used the flag-state squashing model to improve the key generation rate of the unbalanced phase encoding BB84 protocol. The expression for $\widehat{f}(\rho)$ is as follows \cite{tupkary2023using, nickykaihongli2020improving}:
\begin{equation}
	\label{eq:2}
		\min _{\rho \in \mathbf{S}} \widehat{f}(\rho) = \sum_{\tilde{n}=0}^{\infty} p_{\tilde{n}} \min _{\rho_{A B}^{\tilde{n}} \in {\mathbf{S}}_{\tilde{n}}} S\left(Z^R \mid E W \tilde{A} \tilde{B}\right)_{\rho_{A B}^{\tilde{n}}}.\\
\end{equation}
Here, $p_{\tilde{n}}$ denotes the probability of the system being in a state characterized by $\tilde{n}$ photons. $S\left(\cdot \right) $ represents von Neumann entropy. The conditional entropy is minimized over the feasible set of normalized states ${\mathbf{S} }_{\tilde{n}}$ for each output photon number $\tilde{n}$, where $\rho_{A B}^{\tilde{n}} $ is the density matrix. The key information is stored in the standard basis ${\left| j\right\rangle }_{R}$ on a register system R. We then decohere $R$ on this basis, which turns $R$ into a classical register $Z^{R}$. $E$ is a purifying system of $\rho_{A B}^{\tilde{n}} $. $W$ represents the knowledge of the error location of Alice and Bob's data. $\tilde{A}$ and $\tilde{B}$ are classical registers that store Alice’s and Bob’s announcements, respectively. We can see from Eq. (\ref{eq:2}) that $\widehat{f}(\rho)$ can be written in the form of combinations of different photon number pulses. Specifically, $\widehat{f}(\rho)$ represents a minimization problem over the state space $\mathbf{S}$, where each term in the summation corresponds to a different photon number $\tilde{n}$. 

Similarly, the observed information leakage during error correction, denoted as $\operatorname{leak}_{\text{all}}^{\mathrm{EC}}$, can be expressed as a conditional entropy  \cite{winick2018reliable}. This quantity accounts for the information that must be disclosed to correct errors between Alice's and Bob's keys. It can be shown that $\operatorname{leak}_{\text{all}}^{\mathrm{EC}}$ also consists of a combination of pulses with different photon numbers \cite{ winick2018reliable, nickykaihongli2020improving} (See Appendix \ref{appendix:A} for details):
\begin{widetext}
	\begin{equation}
		\label{eq:3}
		\begin{aligned}
			\operatorname{leak}_{\text {all}}^{\mathrm{EC}} & =S\left(Z^R \mid Z^{\bar{B}} \tilde{A} \tilde{B}\right)_{\rho_{A A_{S} B}} \\
			& =S\left(Z^R Z^{\bar{B}} \tilde{A} \tilde{B}\right)_{\rho_{A A_{S} B}} - S\left(Z^{\bar{B}} \tilde{A} \tilde{B}\right)_{\rho_{A A_{S} B}} \\
			& =-\operatorname{Tr}\left(\rho_{Z^R Z^{\bar{B}} \tilde{A} \tilde{B} A A_{S} B} \log _2 \rho_{Z^R Z^{\bar{B}} \tilde{A} \tilde{B} A A_{S} B}\right)+\operatorname{Tr}\left(\rho_{Z^{\bar{B}} \tilde{A} \tilde{B} A A_{S} B} \log _2 \rho_{Z^{\bar{B}} \tilde{A} \tilde{B} A A_{S} B}\right) \\
			& =\sum_{\tilde{n}=0}^{\infty} p_{\tilde{n}}\left[-\operatorname{Tr}\left(\rho_{Z^R Z^{\bar{B}} \tilde{A} \tilde{B} A B}^{\tilde{n}} \log _2 \rho_{Z^R Z^{\bar{B}} \tilde{A} \tilde{B} A B}^{\tilde{n}}\right)+\operatorname{Tr}\left(\rho_{Z^{\bar{B}} \tilde{A} \tilde{B} A B}^{\tilde{n}} \log _2 \rho_{Z^{\bar{B}} \tilde{A} \tilde{B} A B}^{\tilde{n}}\right)\right] \\
			& =\sum_{\tilde{n}=0}^{\infty} p_{\tilde{n}} S\left(Z^R \mid Z^{\bar{B}} \tilde{A} \tilde{B}\right)_{\rho_{A B}^{\tilde{n}}}.
		\end{aligned}
	\end{equation}
\end{widetext}
Here, $Z^{\bar{B}}$ can be viewed as the classical register obtained by measuring on a standardized basis of $\bar{B}$, where $\bar{B}$ is a register that stores Bob's measurement results for a given announcement.. $A_{S}$ is an auxiliary system related to photon number, which is private to Alice. Eq.(\ref{eq:3}) reflects the dependence of the error correction process on the statistical distribution of photon numbers in the quantum states used for key generation. 

Substituting Eq. (\ref{eq:2}), (\ref{eq:3}) into Eq. (\ref{eq:1}) yields
\begin{widetext}
\begin{equation}
	\label{eq:31}
	\begin{aligned}
		R= & p_{\text {pass }} \cdot\left[\sum_{\tilde{n}=0}^{\infty} p_{\tilde{n}} \min _{\rho_{A B}^{\tilde{n}} \in S_{\tilde{n}}} S\left(Z^R \mid E W \tilde{A} \tilde{B}\right)_{\rho_{A B}^{\tilde{n}}}-\sum_{\tilde{n}=0}^{\infty} p_{\tilde{n}} S\left(Z^R \mid Z^{\bar{B}} \tilde{A} \tilde{B}\right)_{\rho_{A B}^{\tilde{n}}}\right] \\
		= & p_{\text {pass }} \cdot\left\{p_{\tilde{n}=0}\left[\min _{\rho_{A B}^{{\tilde{n}}=0} \in S_{{\tilde{n}}=0}} S\left(Z^R \mid E W \tilde{A} \tilde{B}\right)_{\rho_{A B}^{\tilde{n}=0}}-S\left(Z^R \mid Z^{\bar{B}} \tilde{A} \tilde{B}\right)_{\rho_{A B}^{\tilde{n}=0}}\right]\right. \\
		& +p_{\tilde{n}=1}\left[\min _{\rho_{A B}^{{\tilde{n}}=1} \in S_{\tilde{n}=1}} S\left(Z^R \mid E W \tilde{A} \tilde{B}\right)_{\rho_{A B}^{\tilde{n}=1}}-S\left(Z^R \mid Z^{\bar{B}} \tilde{A} \tilde{B}\right)_{\rho_{A B}^{\tilde{n}=1}}\right] \\
		& \left.+\sum_{\tilde{n}=2}^{\infty} p_{\tilde{n}}\left[\min _{\rho_{A B}^{\tilde{n}} \in S_{\tilde{n}}} S\left(Z^R \mid E W \tilde{A} \tilde{B}\right)_{\rho_{A B}^{\tilde{n}}}-S\left(Z^R \mid Z^{\bar{B}} \tilde{A} \tilde{B}\right)_{\rho_{A B}^{\tilde{n}}}\right]\right\}.
		\end{aligned}
\end{equation}
\end{widetext}

We can see from Eq. (\ref{eq:31}) that the SKR is obtained by combining three components: the zero-photon, single-photon, and multi-photon contributions. Each of these components has actual physical significance and should be a non-negative number. Assuming Eve launches the strongest possible attack without violating the principles of quantum mechanics, for polarization-encoded phase-randomized pulses, Eve can perform a PNS attack \cite{norbertlutkenhaus2002quantum}. This implies that the PA part of the third term of the second equal sign in   Eq. (\ref{eq:31}) is theoretically zero because Eve can obtain complete information about the multi-photon states using the PNS attack. In previous calculations, the multi-photon part of the $\operatorname{leak}_{\text{all}}^{\mathrm{EC}}$ was assumed to be greater than zero, i.e,
\begin{widetext}
	\begin{equation}
		\label{eq:35}
		\begin{aligned}
			R= & p_{\text {pass }} \cdot\left\{p_{\tilde{n}=0}\left[\min _{\rho_{A B}^{{\tilde{n}}=0} \in S_{{\tilde{n}}=0}} S\left(Z^R \mid E W \tilde{A} \tilde{B}\right)_{\rho_{A B}^{\tilde{n}=0}}-S\left(Z^R \mid Z^{\bar{B}} \tilde{A} \tilde{B}\right)_{\rho_{A B}^{\tilde{n}=0}}\right]\right. \\
			& +p_{\tilde{n}=1}\left[\min _{\rho_{A B}^{{\tilde{n}}=1} \in S_{\tilde{n}=1}} S\left(Z^R \mid E W \tilde{A} \tilde{B}\right)_{\rho_{A B}^{\tilde{n}=1}}-S\left(Z^R \mid Z^{\bar{B}} \tilde{A} \tilde{B}\right)_{\rho_{A B}^{\tilde{n}=1}}\right] \\
			& \left.+\sum_{\tilde{n}=2}^{\infty}- p_{\tilde{n}}S\left(Z^R \mid Z^{\bar{B}} \tilde{A} \tilde{B}\right)_{\rho_{A B}^{\tilde{n}}}\right\}.
		\end{aligned}
	\end{equation}
\end{widetext}
Therefore, the SKR contribution from the multi-photon part resulted in a negative number, which lacks practical physical sense. 

The combination of different photon number pulses in both $\widehat{f}(\rho)$ and $\operatorname{leak}_{\text{all}}^{\mathrm{EC}}$ underscores the complexity of the key distribution process, where each photon number component contributes differently to the overall security and efficiency of the QKD system. These components must be carefully analyzed and optimized to ensure that the final key remains secure while minimizing the information leakage during error correction. In this paper, we focus on the EC part and propose an improvement to the estimation of this part. This improvement ensures that the third term of Eq. (\ref{eq:31}) is greater than or equal to zero, thereby conferring real physical meaning to it. 
Considering the actual information leakage $\operatorname{leak}_{\text {actual}}^{\mathrm{EC}}$, Eq. (\ref{eq:1}) is changed to
\begin{equation}
	\label{eq:38}
	R = p_{\text {pass }} \cdot \left(\min _{\rho \in \mathbf{S}} \widehat{f}(\rho)-\operatorname{leak}_{\text {actual}}^{\mathrm{EC}}\right),\\
\end{equation}
The actual information leakage $\operatorname{leak}_{\text {actual}}^{\mathrm{EC}}$ is 
\begin{equation}
	\label{eq:36}
		\operatorname{leak}_{\text {actual}}^{\mathrm{EC}} = \sum_{\tilde{n}=0}^{1} p_{\tilde{n}} S\left(Z^R \mid Z^{\bar{B}} \tilde{A} \tilde{B}\right)_{\rho_{A B}^{\tilde{n}}},
\end{equation}
Let 
\begin{equation}
	\label{eq:34}
	\operatorname{leak}_{\text {M}}^{\mathrm{EC}} = \sum_{\tilde{n}=2}^{\infty} p_{\tilde{n}} S\left(Z^R \mid Z^{\bar{B}} \tilde{A} \tilde{B}\right)_{\rho_{A B}^{\tilde{n}}},
\end{equation}
indicating information leakage from the multi-photon part. Then there is
\begin{equation}
	\label{eq:37}
	\operatorname{leak}_{\text {actual}}^{\mathrm{EC}} = \operatorname{leak}_{\text {all}}^{\mathrm{EC}} - \operatorname{leak}_{\text {M}}^{\mathrm{EC}},
\end{equation}

Due to the complexity of von Neumann entropy calculation, we proposed a simple and understandable method to estimate information leakage for the EC part. Owing to the symmetry of DV-QKD protocols, without loss of generality, we only focus on the sequence of Alice. Let $A_0$, $A_1$, $A_M$ represent the set of data bits generated from vacuum, single-photon and multi-photon pulses, respectively, s.t. $A = {A_0} \cup {A_1} \cup {A_M}$, where $A$ is the set containing all data bits. No matter which reconciliation protocol is applied, a collection of data blocks $C= \{ c|c \subset A\}$ will be generated after reconciliation, and each block $c \in C$ is also a set containing several bits from $A$. Ideally, all the blocks from $C$ are linearly independent. In this situation, each block $c$ leaks 1bit information and the estimated value of original information reconciliation leakage $\hat{\operatorname{leak}_{\text {all}}^{\mathrm{EC}}} ={\left| C \right|}/{N}$, where $N$ is the length  of filter code. Let ${C_{M}} = \{ c|(\forall c \in {C}) \wedge (c \subset {A_M})\}$, we have the estimated value of information leakage from the multi-photon part $\hat{\operatorname{leak}_{\text {M}}^{\mathrm{EC}}} = {\left| {{C_M}} \right|}/{N}$. 

Let $D$ represents the set of data block lengths after reconciliation (e.g. $D = \left\{ {8,4,2,1} \right\}$ in the case shown in Fig. \ref{fig:1}) and the sets $C^l, C_{M}^l (l \in D)$ satisfy ${C^l} = \{ c|(\forall c \in C) \wedge (\left| c \right| = l)\}$, $ C_{M}^l = \{ c|(\forall c \in {C_{M}}) \wedge (\left| c \right| = l)\}$. Know that the sequences are shuffled before each pass and the keys are then uniformly distributed in the sequences. It is worth mentioning that this study focuses on an asymptotic SKR estimation (finite analysis will be studied in future work), without considering the influence of statistical fluctuations, which can be regarded as replacement sampling. We have the estimated value of actual information leakage
\begin{equation}
	\label{eq:61}
	\begin{split}
	\hat{\operatorname{leak}_{\text {actual}}^{\mathrm{EC}}} &= \hat{\operatorname{leak}_{\text {all}}^{\mathrm{EC}}} - \hat{\operatorname{leak}_{\text {M}}^{\mathrm{EC}}} \\
	&= {\left| C \right|}/{N} - {\left| {{C_M}} \right|}/{N} \\
    &= \sum\limits_{l \in D} {\frac{\left| {{C^l}} \right|}{N}}  - \sum\limits_{l \in D} {\frac{\left| {C_M^l} \right|}{N}} \\
	&= \sum\limits_{l \in D} {\frac{\left| {{C^l}} \right|}{N}\left[ {1 - {{\left({\Delta _{M}} \right)}^l}} \right]} \\
	&\le \sum\limits_{l \in D} {\frac{\left| {{C^l}} \right|}{N}\left[ {\min \left( {1 - {{\left(\Delta _{M}^{\min } \right)}^l},1} \right)} \right]}, \\
	\end{split}
\end{equation}
where $\Delta _{M}$ represents the proportion of the count rate of the multi-photon pulses to the signal pulses. $\Delta _{M}^{\min }$ is the lower limit of $\Delta _{M}$.

We eventually get the formula for calculating the improved $R$:
\begin{widetext}
\begin{equation}
    \label{eq:62}
        R = q {Q_\mu }\max \left\{ \min _{\rho \in \mathbf{S}} \hat{f}(\rho)- \sum\limits_{l \in D} \left( \frac{|C^l|}{N} \right) \left[ \min \left( 1 - \left( \Delta _{M}^{\min} \right)^l, 1 \right) \right], 0\right\}.
\end{equation}
\end{widetext}

Here, $q$ is the probability of successful screening. $Q_\mu$ is the signal state response rate. We can see from Eq. (\ref{eq:62}) that the critical parameter is $\Delta _{M}^{\min }$. To this end, we give the methods for calculating $\Delta _{M}^{\min }$ for two typical DV-QKD protocols based on the GLLP theory, that is, decoy-BB84 and MDI protocols. In addition, we can clearly see from Eq. (\ref{eq:62}) that the smaller $l$ is, the better our approach performs. However, in a commonly used non-interactive reconciliation protocol, a larger $l$ is usually applied to achieve a higher $f$. Therefore, we believe that these reconciliation protocols are unable to gain obvious SKR improvement from our approach. In contrast, a collection of data blocks with different block lengths whose value can be as low as 1, will be obtained after Cascade reconciliation. We thus conclude that a greater SKR improvement can be obtained when using Cascade reconciliation. In addition, considering that the $f$ of Cascade has been rather high, we believe that Cascade is the best-fit reconciliation protocol.
      
\subsection{Decoy-BB84 protocol}
The most special type of protocol in the decoy-state protocol is called the weak+vacuum decoy-state protocol, in which the average number of photons in the optical pulse satisfies $\nu_2\ll \nu_1<\mu$ (where $\mu$ is the average number of photons in the signal state, and $\nu_1$ and $\nu_2$ are the average number of photons in the two decoy states.). Known that  $\Delta _{M}^{\min } = 1 - \Delta _0^{\max } - \Delta _1^{\max }$ in a decoy-BB84 protocol, where $\Delta _{i}^{\max }$ is the upper limit of $\Delta _{i}$ ($i=0, 1$), we just need to focus on the calculation of $\Delta _0^{\max }$ and $\Delta _1^{\max }$. ${\Delta _i} = \frac{{{Q_i}}}{{{Q_\mu }}}$ and $Q_i$ represents the count rate of the pulse whose photon number is $i (i = 0,1,2,...,n)$. By using the decoy-state method, we have
\begin{equation}
	\label{eq:15}
	\begin{split}
	{Q_{{\nu _1}}} &= \sum\limits_{i = 0}^\infty  {{Y_i}} \frac{{\nu _1^i}}{{i!}}{e^{ - {\nu _1}}}, \\
	{Q_{{\nu _2}}} &= \sum\limits_{i = 0}^\infty  {{Y_i}} \frac{{\nu _2^i}}{{i!}}{e^{ - {\nu _2}}}, \\
	\end{split}
\end{equation}
where $Q_{\nu _i}$ represents the count rate of the pulse whose average photon number is $\nu _j (j = 1,2)$. $Y_{i}$ is the probability of Bob's detector response when Alice sends a photon pulse containing $i$ photons.

Based on Eq. (\ref{eq:15}), we have
\begin{equation}
	\label{eq:16}
    \begin{split}
	{Q_{{\nu_1}}}{e^{{\nu_1}}} - {Q_{{\nu_2}}}{e^{{\nu_2}}} &= {Y_1}({\nu_1} - {\nu_2}) + \sum\limits_{i=2}^\infty  {\frac{{{Y_i}}}{{i!}}(\nu_1^i  - \nu_2^i)} \\
    &\ge {Y_1}({\nu_1} - {\nu_2}).
    \end{split}
\end{equation}

Then the upper bounds of $Y_1$ and ${\Delta _1}$ can be obtained as follow:
\begin{equation}
	\label{eq:17}
	\begin{split}
	{Y_1} &\le \frac{{{Q_{{v_1}}}{e^{{\nu _1}}} - {Q_{{\nu_2}}}{e^{{\nu _2}}}}}{{{\nu _1} - {\nu _2}}}, \\
	{\Delta _1} &= \frac{{{Q_1}}}{{{Q_\mu }}} = \frac{{{Y_1}\mu {e^{ - \mu }}}}{{{Q_\mu }}} \\
    &\le \frac{{({Q_{{\nu_1}}}{e^{{\nu _1}}} - {Q_{{\nu_2}}}{e^{{\nu _2}}})\mu{e^{ -\mu}}}}{{({\nu_1} - {\nu_2}){Q_\mu}}}. \\
	\end{split}
\end{equation}

For $Y_0$ and ${\Delta _0}$, according to Eq. (\ref{eq:15}), we have
\begin{equation}
	\label{eq:18}
    \begin{split}
	{Q_{{\nu_2}}}{e^{{\nu_2}}} &= {Y_0} + {Y_1}{\nu_2} + \sum\limits_{i = 2}^\infty  {{Y_i}} \frac{{\nu_2^i}}{{i!}} \\
    &\ge {Y_0} + {Y_1}{\nu_2},
    \end{split}
\end{equation}

thus the upper bounds of $Y_0$ and ${\Delta _0}$ can be obtained as
\begin{equation}
	\label{eq:19}
	\begin{split}
	{Y_0} &\le {Q_{{\nu_2}}}{e^{{\nu_2}}} - {Y_1}{\nu_2} \\
	&\le {Q_{{\nu_2}}}{e^{{\nu_2}}} - Y_1^{\min }{\nu_2}, \\
	{\Delta _0} &= \frac{{{Q_0}}}{{{Q_\mu}}} = \frac{{{Y_0}{e^{ -\mu}}}}{{{Q_\mu}}} \\
	&\le \frac{{({Q_{{\nu_2}}}{e^{{\nu_2}}} - Y_1^{\min }{\nu_2}){e^{ -\mu}}}}{{{Q_\mu}}}. \\
	\end{split}
\end{equation}

According to Eq. (\ref{eq:17}) and (\ref{eq:19}), we can calculate the value of $\Delta _{M}^{\min }$, where the formula of $Y_1^{\min }$ and ${Q_{\mu}}$ have been given in Ref. \cite{ma2005practical}. 
\subsection{MDI protocol}
Similar to the decoy-BB84 protocol, $\Delta _{MM}^{\min } = 1 - \Delta _{00}^{\max } - \Delta _{01}^{\max } - \Delta _{10}^{\max } - \Delta _{11}^{\max }$, we need to compute $\Delta _{00}^{\max }$, $\Delta _{01}^{\max }$, $\Delta _{10}^{\max }$ and $\Delta _{11}^{\max }$. ${\Delta _{ij}} = \frac{{{Q_{ij}}}}{{{Q_{\mu\mu }}}}$, where $Q_{\mu\mu}$ is the signal state response rate, and $Q_{ij}$ represents the count rate of pulses with photon numbers $i (i=0,1,2,..., n) $ and $j (j=0,1,2,..., n) $ from Alice and Bob, respectively. $\Delta _{MM}^{\min }$ is the lower limit of $\Delta _{MM}$, and $\Delta _{ij}^{\max }$ is the upper limit of $\Delta _{ij}(i,j =0,1)$. We have
\begin{equation}
	\label{eq:20}
	\begin{aligned}
	& \Delta_{00}=\frac{Q_{00}}{Q_{\mu \mu}}=\frac{Y_{00} e^{-2 \mu}}{Q_{\mu \mu}} \leq \frac{Y_{00}^{\max } e^{-2 \mu}}{Q_{\mu \mu}} = \Delta_{00}^{\max }, \\
	& \Delta_{01}=\frac{Q_{01}}{Q_{\mu \mu}}=\frac{Y_{01} \mu e^{-2 \mu}}{Q_{\mu \mu}} \leq \frac{Y_{01}^{\max } \mu e^{-2 \mu}}{Q_{\mu \mu}} = \Delta_{01}^{\max }, \\
	& \Delta_{10}=\frac{Q_{10}}{Q_{\mu \mu}}=\frac{Y_{10} \mu e^{-2 \mu}}{Q_{\mu \mu}} \leq \frac{Y_{10}^{\max } \mu e^{-2 \mu}}{Q_{\mu \mu}} = \Delta_{10}^{\max }, \\
	& \Delta_{11}=\frac{Q_{11}}{Q_{\mu \mu}}=\frac{Y_{11} \mu^2 e^{-2 \mu}}{Q_{\mu \mu}} \leq \frac{Y_{11}^{\max } \mu^2 e^{-2 \mu}}{Q_{\mu \mu}} = \Delta_{11}^{\max }, 
	\end{aligned}
\end{equation}
where $Y_{ij}^{\max }$ is the upper limit of $Y_{ij}(i,j=0,1)$. From the decoy state analysis and the following linear constraint, 
\begin{equation}
	\label{eq:21}
	\begin{aligned}
	Q_{\nu_{A_i} \nu_{B_j}} & \leq \sum_{n_A \leq N} \sum_{n_B \leq N} p_{\nu_{A_i}}\left(n_A\right) p_{\nu_{B_j}}\left(n_B\right) Y_{n_A, n_B} \\
	& \quad +\left(1-\sum_{n_A \leq N} \sum_{n_B \leq N} p_{\nu_{A_i}}\left(n_A\right) p_{\nu_{B_j}}\left(n_B\right)\right), \\
	Q_{\nu_{A_i} \nu_{B_j}} & \geq \sum_{n_A \leq N} \sum_{n_B \leq N} p_{\nu_{A_i}}\left(n_A\right) p_{\nu_{B_j}}\left(n_B\right) Y_{n_A, n_B} .
	\end{aligned}
\end{equation}
we can solve for the values of $Y _{00}^{\max }$, $Y _{01}^{\max }$, $Y _{10}^{\max }$ and $Y _{11}^{\max }$. $p_{\nu_{A_i}}\left(n_A\right)$ ($p_{\nu_{B_j}}\left(n_B\right)$) represents the probability of photon number $n_A$($n_B$) when Alice's(Bob's) decoy-state intensity is $\nu_{A_i}$($\nu_{B_j}$), and follows the Poisson distribution:
\begin{equation}
	\label{eq:22}
	\begin{aligned}
	p_{\nu_{A i}}\left(n_A\right) & =\frac{\nu_{A i}{ }^{n_A}}{n_{A}!} e^{-\nu_{A i}}, \\
	p_{\nu_{B j}}\left(n_B\right) & =\frac{\nu_{B j}{ }^{n_B}}{n_{B}!} e^{-\nu_{B j}},
	\end{aligned}
\end{equation}

According to Eq. (\ref{eq:20}), we can calculate the value of $\Delta _{MM}^{\min }$, where the calculation method of ${Q_{\mu \mu}}$ have been given in Ref. \cite{wang2021numerical}. 

\section{Simulation and Discussion}
\label{sec:4}
\subsection{Simulation result}
We evaluate the performance improvement of our new approach over the original GLLP formula through simulations under practical experimental parameters. In addition, we applied Cascade as the reconciliation protocol in our simulations, and its reconciliation efficiency $f$ is set to the optimal value 1. For simplicity, only information leakage with a block length of 1 was calculated, i.e., $l$ in Eq. (\ref{eq:62}) is set to 1.

\begin{figure}
	\includegraphics[width=0.5\textwidth]{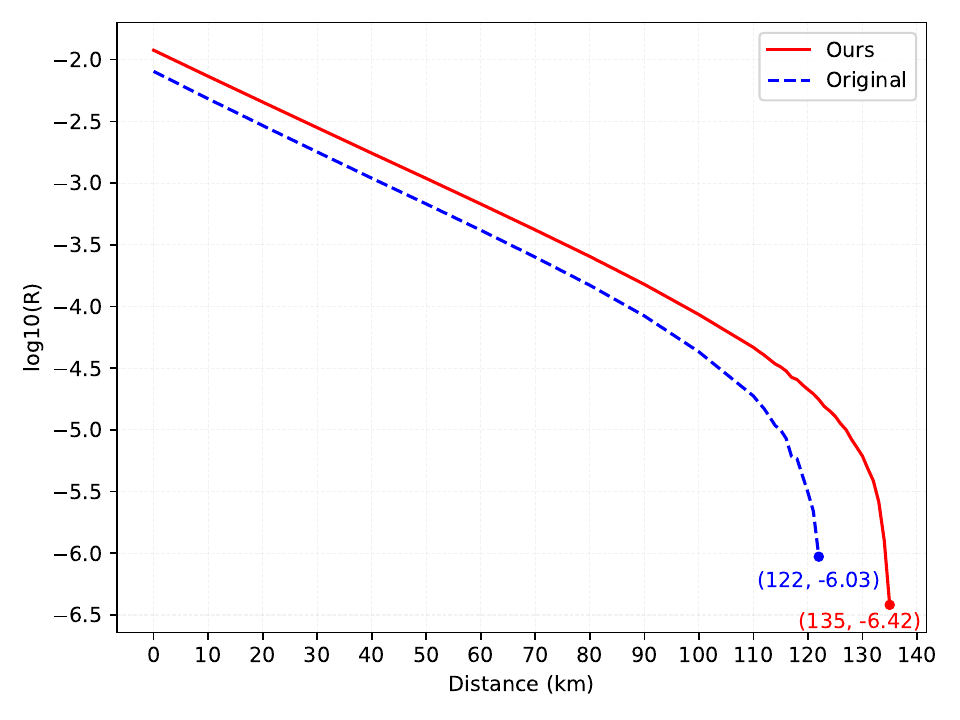}
	\caption{\label{fig:2}SKR (taked the logarithm based on 10) vs transmission distance with our improved and the original approach in a decoy-BB84 protocol \cite{tupkary2023using}. The key simulation parameters are: $\mu_1$, $\mu_2$, $\mu_3$: average photon number for signal, decoy and vacuum pulses, are 0.4, 0.1, and 0.0007, respectively; $\eta$: the channel loss, is 0.20 dB/km; $d$: the dark count rate of the detector, is $10^{-5}$; $\eta_{d}$: detector efficiency, is 0.2; $e_{det}$: misalignment, defined as single-photon error, is 0.033.}
\end{figure}

\begin{figure}
	\includegraphics[width=0.5\textwidth]{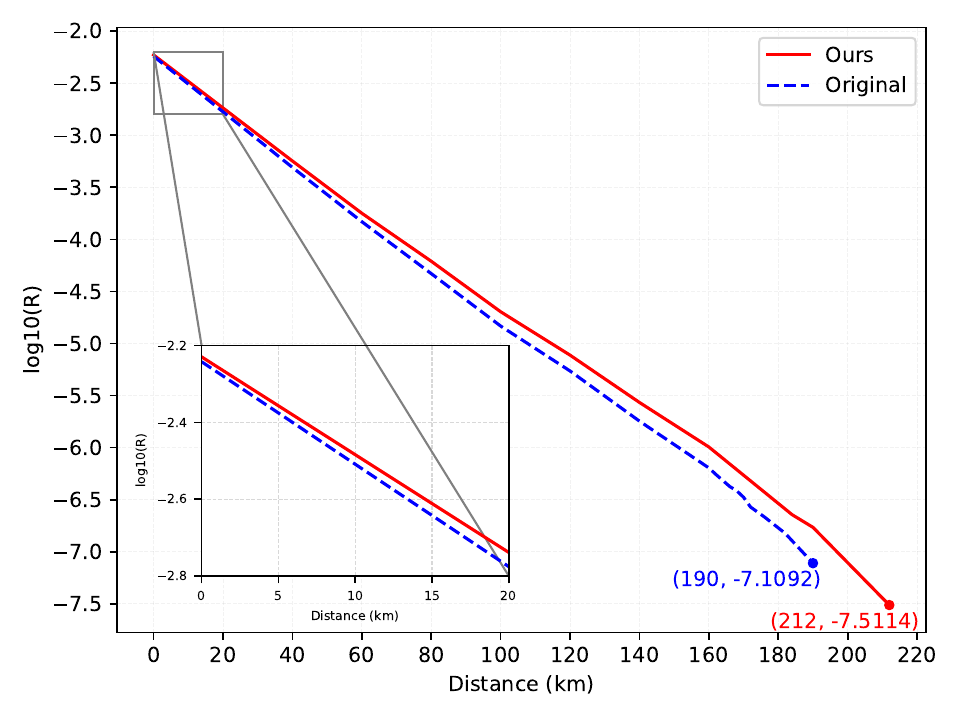}
	\caption{\label{fig:3}SKR (taked the logarithm based on 10) vs transmission distance with our improved and the original approach in a MDI protocol \cite{wang2021numerical, lo2012measurementdeviceindependent}. The key simulation parameters are: $\mu_1$, $\mu_2$, $\mu_3$: average photon number for signal, decoy and vacuum pulses, are 0.4, 0.1, and 0.0007, respectively; $\eta$: the channel loss, is 0.20 dB/km; $d$: the dark count rate of the detector, is $10^{-6}$; $\theta_{A}$: misalignment angle between Alice and Charlie; $\theta_{B}$: misalignment angle between Bob and Charlie. The misalignment of Alice and Bob is set to $\theta_{A}=0.091$ and $\theta_{B}=-0.091$ (equivalent to approximately 0.033 misalignment between Alice and Bob). For simplicity, the detector efficiency is included in the total loss.}
\end{figure}

As shown in Fig. (\ref{fig:2}) and Fig. (\ref{fig:3}), the simulation results show that our SKRs surpass the original results at any distance, where the sub-graph in Fig. (\ref{fig:3}) shows local amplification of SKRs from 0KM to  20KM. In addition, the results of decoy-BB84 and MDI protocols show an increase in transmission distance of about 13KM and 22KM, respectively. Moreover, as shown in Table \ref{tab:3}, the farther the transmission distance, the more the value of SKR increases. For example, the SKR enhancement rates of the two protocols are 100.40\% and 37.17\% at 100KM, respectively. It is worth noting that the improvement in SKRs with the MDI protocol, compared to the decoy-BB84 protocol, is relatively small. This is mainly because the MDI protocol offers better security and higher tolerance when dealing with multi-photon pulses, which reduces the impact of the multi-photon effect on SKRs\cite{gisin2002quantum, lo2012measurementdeviceindependent}. The results indicate that the SKRs of our method has some increase over the original method.

\begin{table}
	\caption{\label{tab:3}SKR growth rate for the decoy-BB84 and MDI protocol.}
	\begin{ruledtabular}
	\begin{tabular}{ccccccc}
		\multicolumn{1}{c}{} & \multicolumn{6}{c}{Transmission distance(km)} \\
		\cline{2-7}
		\multicolumn{1}{c}{} & 0 & 20 & 40 & 60 & 80 & 100 \\
		\hline
		decoy-BB84 & 48.36\% & 55.04\% & 59.48\% & 63.14\% & 71.03\% & 100.40\%\\
		MDI & 3.27\% & 14.00\% & 18.50\% & 20.56\% & 31.45\% & 37.17\% \\
		\end{tabular}
	\end{ruledtabular}
\end{table}

\subsection{Discussion}
Encouraged by the promising performance of our approach, we predict that it can significantly advance the development of QKD systems, particularly in the realm of QKD post-processing research. It is crucial to highlight that our method adheres to the Shannon theorem. The required information leakage of $L_{\text{all}} = N h(e)$ ensures reliable reconciliation; however, a portion of $L_{\text{all}}$ is already known to Eve before reconciliation. As a result, the actual information leakage $L_{\text{actual}}$ for Eve is less than $N h(e)$.

Given this context, we suggest that Shannon entropy is no longer a suitable metric to describe the minimum information leakage during reconciliation, necessitating the development of a new indicator. This new metric should more accurately reflect the true information leakage in the presence of known information pre-reconciliation.

Our approach shows significant potential in enhancing the SKR at any distance. However, Eq. (\ref{eq:62}) requires further modification for application in the finite-size regime. Finite-size effects also impact the information leakage in our new approach, indicating that a large input size may be necessary for practical implementation. Additionally, a better bound for $\Delta_{M}^{\min}$ might be achieved through alternative derivations, which would further contribute to the improvement of the SKR. 

\section{Concluding Remarks}
\label{sec:5}
In this study, we propose a novel approach to reduce the information leakage during reconciliation by specifically considering the effect of the quantum part on the post-processing part. By eliminating the information leakage caused by multi-photon pulses, our proposed approach surpasses the previous theoretical limits for information leakage during reconciliation. For both the decoy-BB84 and MDI protocols, simulation results demonstrate that our approach can significantly increase the key rate at any distance. Interestingly, the improvement percentage grows with the increase in distance, and our approach also extends the maximum transmission distance. Notably, our method is applicable to all DV-QKD protocols based on the GLLP theory.

Moreover, we believe that the central idea of our approach may have significant implications for Continuous-Variable (CV) QKD protocols as well. Our core concept suggests that quantum and classical signal processing are not entirely independent. Therefore, the impact of quantum pulses on classical data bits should be considered when estimating information leakage through the classical channel. Once some information leakage $\operatorname{leak}_{\text {all}}^{M}$ has been acquired by Eve before reconciliation, regardless of the type of QKD protocol, $\operatorname{leak}_{\text {all}}^{M}$ should not be disregarded when calculating the SKR. This analysis leads us to an important implication for CV-QKD systems: better SKRs can be achieved in CV-QKD systems if similar interdependencies exist. By acknowledging the interdependence between quantum pulses and classical data bits, we can enhance the accuracy of information leakage estimations and improve the overall performance of QKD protocols.
\begin{acknowledgments}
This work is supported by National Natural Science Foundation of China (Grant number: 62071151), and Innovation Program for Quantum Science and Technology (Grant No. 2021ZD0300701). Special thanks goes to Prof. Xiong-Feng Ma for the valuable discussions. We gratefully acknowledge the constructive feedback provided by the editor and reviewers, whose insights helped to enhance the quality of this paper.
\end{acknowledgments}

H.-K. M. and B. Y. contributed equally to this work.

\onecolumngrid
\appendix
\section{PROOF OF DECOMPOSING THE ERROR CORRECTION TERM}
\label{appendix:A}
In sec. (\ref{sec:3}), Eq. (\ref{eq:3}), $\operatorname{leak}_{\text{all}}^{\mathrm{EC}}$ can be written in the following form,
	\begin{equation}
		\label{eq:4}
		\begin{aligned}
			\operatorname{leak}_{\text {all}}^{\mathrm{EC}} & =S\left(Z^R \mid Z^{\bar{B}} \tilde{A} \tilde{B}\right)_{\rho_{A A_{S} B}} \\
			& =S\left(Z^R Z^{\bar{B}} \tilde{A} \tilde{B}\right)_{\rho_{A A_{S} B}} - S\left(Z^{\bar{B}} \tilde{A} \tilde{B}\right)_{\rho_{A A_{S} B}},
		\end{aligned}	
	\end{equation}
where $A_{S}$ is an ancillary system related to photon number, which is private to Alice. The calculation method for the first term of Eq. (\ref{eq:4}) is as follows
\begin{equation}
	\label{eq:5}
		S\left(Z^R Z^{\bar{B}} \tilde{A} \tilde{B}\right)_{\rho_{A A_{S} B}} =-\operatorname{Tr}\left(\rho_{Z^R Z^{\tilde{B}} \tilde{A} \tilde{B} A A_{S} B} \log _2 \rho_{Z^R Z^{\tilde{B}} \tilde{A} \tilde{B} A A_{S} B}\right),
\end{equation}
since
\begin{equation}
	\label{eq:6}
		\rho_{Z^R Z^{\tilde{B}} \tilde{A} \tilde{B} A A_{S} B} =\sum _ { \tilde { n } = 0 } ^ { \infty } p _ { \tilde { n } } |\tilde{n}\rangle\left\langle\left.\tilde{n}\right|_{A_S}\right. \otimes \rho _ { Z ^ { R } Z ^ { \bar { B } } \tilde { A } \tilde { B } A B} ^ { \tilde {n} } ,
\end{equation} 
then
\begin{equation}
	\label{eq:12}
		\log _2 \rho_{Z^R Z^{\tilde{B}} \tilde{A} \tilde{B} A A_{S} B} =\sum _ { \tilde { n } = 0 } ^ { \infty }  |\tilde{n}\rangle\left\langle\left.\tilde{n}\right|_{A_S}\right. \otimes \left[\left(\log_2 p_{\tilde{n}}\right) 1_{Z^R Z^{\bar { B }} \tilde{A} \tilde{B} A B}+\log_2 \rho_{Z^R Z^{\bar { B }} \tilde{A} \tilde{B} A B}^{\tilde{n}}\right].
\end{equation} 

Eq. (\ref{eq:5}) can be obtained as
\begin{equation}
	\label{eq:7}
	\begin{aligned}
		S\left(Z^R Z^{\bar{B}} \tilde{A} \tilde{B}\right)_{\rho_{A A_{S} B}} & = -\operatorname{Tr}\left(\sum _ { \tilde { n } = 0 } ^ { \infty } p _ { \tilde { n } } |\tilde{n}\rangle\left\langle\left.\tilde{n}\right|_{A_S}\right. \otimes \rho _ { Z ^ { R } Z ^ { \bar { B } } \tilde { A } \tilde { B } A B} ^ { \tilde {n} } \right) \\
		& \quad \quad\cdot\left(\sum_{\tilde{m}=0}^{\infty}|\tilde{m}\rangle\left\langle\left.\tilde{m}\right|_{A_S}\right. \otimes \left[\left(\log_2 p_{\tilde{m}}\right) 1_{Z^R Z^{\bar { B }} \tilde{A} \tilde{B}  A B}+\log_2 \rho_{Z^R Z^{\bar { B }} \tilde{A} \tilde{B}  A B }^{\tilde{m}}\right]\right) \\
		& = -\sum _ { \tilde { n } = 0 } ^ { \infty }\operatorname{Tr}\left( p _ { \tilde { n } } |\tilde{n}\rangle\left\langle\left.\tilde{n}\right|_{A_S}\right. \otimes \rho _ { Z ^ { R } Z ^ { \bar { B } } \tilde { A } \tilde { B } A B} ^ { \tilde {n} } \right) \\
		& \quad \quad\cdot\left(|\tilde{n}\rangle\left\langle\left.\tilde{n}\right|_{A_S}\right. \otimes \left[\left(\log_2 p_{\tilde{n}}\right) 1_{Z ^ { R } Z ^ { \bar { B } } \tilde { A } \tilde { B } A B}+\log_2 \rho_{Z ^ { R } Z ^ { \bar { B } } \tilde { A } \tilde { B } A B}^{\tilde{n}}\right]\right) \\
		& = -\sum _ { \tilde { n } = 0 } ^ { \infty }[\operatorname{Tr}\left( p _ { \tilde { n } } |\tilde{n}\rangle\left\langle\left.\tilde{n}\right|_{A_S}\right. \otimes \rho _ {Z ^ { R } Z ^ { \bar { B } } \tilde { A } \tilde { B } A B} ^ { \tilde {n} } \right) \left(|\tilde{n}\rangle\left\langle\left.\tilde{n}\right|_{A_S}\right. \otimes \left(\log_2 p_{\tilde{n}}\right) 1_{Z ^ { R } Z ^ { \bar { B } } \tilde { A } \tilde { B } A B}\right) \\
		& \quad \quad + \operatorname{Tr}\left( p _ { \tilde { n } } |\tilde{n}\rangle\left\langle\left.\tilde{n}\right|_{A_S}\right. \otimes \rho _ {Z ^ { R } Z ^ { \bar { B } } \tilde { A } \tilde { B } A B} ^ { \tilde {n} } \right) \left(|\tilde{n}\rangle\left\langle\left.\tilde{n}\right|_{A_S}\right. \otimes \log_2 \rho_{Z ^ { R } Z ^ { \bar { B } } \tilde { A } \tilde { B } A B}^{\tilde{n}}\right)],
	\end{aligned}
\end{equation}
because
\begin{equation}
	\label{eq:8}
	\begin{aligned}
 		\operatorname{Tr}(|\tilde{n}\rangle\left\langle\left.\tilde{n}\right|_{A_S}\right.)&=1, \\
  		\operatorname{Tr}(\rho _ {Z ^ { R } Z ^ { \bar { B } } \tilde { A } \tilde { B } A B} ^ { \tilde {n} })&=1, 
	\end{aligned}	
\end{equation}
we have
	\begin{equation}
		\label{eq:9}
		\begin{aligned}
			S\left(Z^R Z^{\bar{B}} \tilde{A} \tilde{B}\right)_{\rho_{A A_{S} B}}  & = - \sum _ { \tilde { n } = 0 } ^ { \infty } \left[ p _ { \tilde { n } } \left(\log_2 p_{\tilde{n}}\right) + p _ { \tilde { n } } \operatorname{Tr} \left( \rho _ {Z ^ { R } Z ^ { \bar { B } } \tilde { A } \tilde { B } A B} ^ { \tilde {n} } \log_2 \rho_{Z ^ { R } Z ^ { \bar { B } } \tilde { A } \tilde { B } A B}^{\tilde{n}} \right)\right]\\
			& = - \sum _ { \tilde { n } = 0 } ^ { \infty } p _ { \tilde { n } }\left[  \left(\log_2 p_{\tilde{n}}\right) +  \operatorname{Tr} \left( \rho _ {Z ^ { R } Z ^ { \bar { B } } \tilde { A } \tilde { B } A B} ^ { \tilde {n} } \log_2 \rho_{Z ^ { R } Z ^ { \bar { B } } \tilde { A } \tilde { B } A B}^{\tilde{n}} \right)\right].
		\end{aligned}
	\end{equation}

 Similarly, the calculation result of the second term in Eq. (\ref{eq:4}) is
\begin{equation}
	\label{eq:10}
	\begin{aligned}
		S\left(Z^{\bar{B}} \tilde{A} \tilde{B}\right)_{\rho_{A A_{S} B}} & =-\operatorname{Tr}\left(\rho_{ Z^{\tilde{B}} \tilde{A} \tilde{B} A A_{S} B} \log _2 \rho_{ Z^{\tilde{B}} \tilde{A} \tilde{B} A A_{S} B}\right) \\
		& = - \sum _ { \tilde { n } = 0 } ^ { \infty } p _ { \tilde { n } }[  \left(\log_2 p_{\tilde{n}}\right) +  \operatorname{Tr} \left( \rho _ {  Z ^ { \bar { B } } \tilde { A } \tilde { B } A B} ^ { \tilde {n} } \log_2 \rho_{ Z^{\bar { B }} \tilde{A} \tilde{B} A B}^{\tilde{n}} \right)].
	\end{aligned}
\end{equation}

Substituting Eq. (\ref{eq:9}, \ref{eq:10}) into Eq. (\ref{eq:4}) yields
\begin{equation}
	\label{eq:11}
	\begin{aligned}
		\operatorname{leak}_{\text {all}}^{\mathrm{EC}} & =S\left(Z^R \mid Z^{\bar{B}} \tilde{A} \tilde{B}\right)_{\rho_{A A_{S} B}} \\
		& =\sum_{\tilde{n}=0}^{\infty} p_{\tilde{n}}\left[-\operatorname{Tr}\left(\rho_{Z^R Z^{\bar{B}} \tilde{A} \tilde{B} A B}^{\tilde{n}} \log _2 \rho_{Z^R Z^{\bar{B}} \tilde{A} \tilde{B} A B}^{\tilde{n}}\right)+\operatorname{Tr}\left(\rho_{Z^{\bar{B}} \tilde{A} \tilde{B} A B}^{\tilde{n}} \log _2 \rho_{Z^{\bar{B}} \tilde{A} \tilde{B} A B}^{\tilde{n}}\right)\right] \\
		& =\sum_{\tilde{n}=0}^{\infty} p_{\tilde{n}} S\left(Z^R \mid Z^{\bar{B}} \tilde{A} \tilde{B}\right)_{\rho_{A B}^{\tilde{n}}}.
	\end{aligned}
\end{equation}
\twocolumngrid

\bibliography{ref}

\end{document}